# Giant Magnetoresistance by Exchange Springs in $DyFe_2$/$YFe_2$ Superlattices

S. N. Gordeev[1], J-M. L. Beaujour[1], G. J. Bowden[1], P. A. J. de Groot[1], B. D. Rainford[1],
R. C. C. Ward[2], M. R. Wells[2], and A. G. M. Jansen[3]

[1]*Department of Physics and Astronomy, University of Southampton, SO17 1BJ, UK*
[2]*Clarendon Laboratory, Oxford University OX1 3PU, UK*
[3]*Grenoble High Magnetic Field Laboratory, MPIF, F-38042 Grenoble Cedex 9, France*

Magnetization and magnetoresistance measurements are reported for antiferromagnetically coupled $DyFe_2$/$YFe_2$ multilayers in fields up to 23 T. We demonstrate that the formation of short exchange springs (~ 20 Å) in the magnetically soft $YFe_2$ layers results in a giant magneto-resistance as high as 32% in the spring region. It is shown that both the magnitude of the effect, and its dependence on magnetic field, are in good agreement with the theory of Levy and Zhang for giant magnetoresistance due to domain wall like structures.

PACS: 73.61.At, 75.60Ch, 75.70.Cn, 75.70.Pa

Giant magnetoresistance (GMR) materials challenge our understanding of spin-polarised electron transport and have many important technical applications (sensors, magnetic read heads etc. [1]). GMR is observed in a variety of magnetic systems including multilayers [2], spin valves [3] and granular materials [4]. A typical GMR multilayer consists of ferromagnetic layers, separated by *nonmagnetic* spacer layers. The effect occurs because of a large difference in the scattering rates of spin-up and spin-down electrons in the magnetic layers. The resistance of a given GMR structure is maximum when magnetic moments in the consecutive magnetic layers are antiparallel and minimum when they are parallel. When a magnetic field is applied, the directions of magnetization in consecutive layers can be switched, resulting in a dramatic change in the resistance of the structure as a whole.

It has been argued that a domain wall in a ferromagnet should give rise to a GMR-like magnetoresistance [5, 6]. In particular, Viret et al. [5] suggested that the inablity of an electron spin to track the reorientation of magnetization in domain walls, could give rise to GMR. Subsequently, a more complete, quantum mechanical, treatment of domain wall GMR was put forward by Levy and Zhang [7] describing both the case with currents in the plane (CIP) and with currents perpendicular to the plane (CPP). According to this model, electron spin mis-tracking causes a mixing of states with opposite spins resulting in GMR due to spin-dependent scattering by impurity centres within the domain walls. In particular, the magnitude of the effect should increase rapidly with decreasing domain wall width ($\delta_w$).

Experimentally, GMR has been witnessed in striped domain walls, in magnetically prepared thin films of Ni and Co [6] and FePd [8]. However, in practice, considerable difficulties were experienced. In the first place, the GMR effect is small (< 1.5%), because of the low density of domain walls and their comparatively large widths $\delta_w$ (>75 Å) [7-9]. Secondly, it is difficult to separate the normal anisotropic magnetoresistance (AMR) from that of GMR. These difficulties can be overcome using artificially tailored domain walls known as "magnetic exchange springs". In this Letter, it is argued that exchange spring multilayers are ideal systems for GMR studies. Firstly, the 'artificial domain walls' occupy a significant fraction of the sample volume (up to ~50%). Secondly, their widths $\delta_w$ can be varied over a wide range either by changing the thickness of the magnetically soft layers, or the strength of the applied magnetic field, or both. Thirdly magnetization measurements can be used to obtain values for $\delta_w$ of the exchange springs. In this context, it should be noted that Mibu et al. [10] have studied the effect of exchange springs in the *ferromagnetically coupled* bilayer system SmCo/NiFe. However, the measured magnetoresistance (MR) was small (~1.5%) and dominated by AMR. The contribution from GMR, estimated at ~ 0.1%, is weak because the SmCo/NiFe system cannot exhibit magnetic exchange springs shorter than ~ 500 Å. It is known [11] that $\delta_w$ decreases with increasing the applied field, $B_{app}$. However, in ferromagnetically coupled multilayers exchange springs disappear when $B_{app}$ exceeds the coercive field, $B_C$, of the magnetically hard layers.

In this Letter, we report the first observation of significant exchange spring driven GMR. We used a $DyFe_2$/$YFe_2$ multilayer system, which allows very short exchange springs to be set up, as a direct result of the *antiferromagnetic* coupling between the $DyFe_2$ and $YFe_2$ layers. Our calculations show that exchange springs as short as 20 Å can be set up in the $YFe_2$ layers. In addition, it is shown that both AMR and GMR effects can be clearly separated. Experimentally, the GMR results are found to be in good agreement with the model of Levy and Zhang [7], both in magnitude and the functional dependence on the domain wall width $\delta_w$.

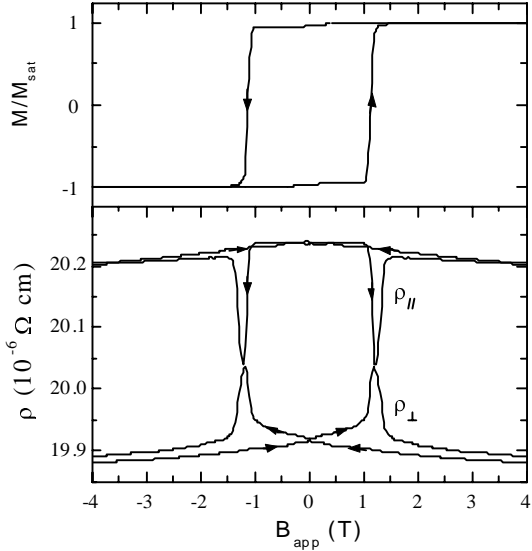

FIG. 1. The field dependent magnetization (a) and resistivity curves ($\rho_\parallel$ and $\rho_\perp$), for currents both parallel and perpendicular to the magnetic field $B_{app}$ (b), in a 4000Å thick $DyFe_2$ film at 100 K.

The Laves phase $DyFe_2/YFe_2$ superlattices (110) were grown by molecular beam epitaxy technique as described elsewhere [12, 13]. The samples were grown on epi prepared $(11\bar{2}0)$ sapphire substrates with a 500 Å (110) Nb buffer and a 30 Å seed layer of Fe. The magnetic measurements were made using a 12 T vibrating sample magnetometer. Transport studies were performed for 0.2 mA currents in plane (CIP), both parallel to and perpendicular to $B_{app}$ using a standard four point method. All resistivity results presented here were corrected for the contribution from the buffer layer. For all measurements $B_{app}$ // [001] (the easy magnetization direction).

The magnetization of a 4000 Å $DyFe_2$ film, together with the MR curves, for both the current parallel ($\rho_\parallel$) and perpendicular ($\rho_\perp$) to $B_{app}$ are presented in Fig 1. Both resistivity curves are found to be hysteretic and there are strong peaks (troughs) at the coercive field of $B_C = 1.2$ T. Note that the two MR curves show effects with opposite sign, implying that the effect is due to AMR. This reciprocity, in the vicinity of $B_C$, is typical of ferromagnetic materials [6].

The properties of the $DyFe_2/YFe_2$ superlattices are characterised by (i) dominant ferromagnetic Fe-Fe exchange (~600 T), (ii) a weaker antiferromagnetic Dy-Fe exchange (~100 T), and (iii) a Dy crystal field anisotropy (~10-100 T), which determines the direction of easy magnetization in the magnetically-hard $DyFe_2$ layers. Our previous magnetization studies [14] of $DyFe_2/YFe_2$ superlattices have shown that exchange springs, set up within the soft $YFe_2$ layers, play an important role in the determination of the magnetization of the multilayer film. For fields less than a critical bending field, $B_B$, all the Fe magnetic moments in both $YFe_2$ and $DyFe_2$ layers (magnetic moment of Y is small) are aligned parallel to each other, due to strong Fe-Fe magnetic exchange (see insets to Fig. 2). For $B_{app} > B_B$, the magnetic moments within $YFe_2$ layers start to rotate, forming exchange springs which are pinned only at the edges by the neighbouring $DyFe_2$ layers. This results in a reversible increase of the magnetization above $B_B$ (e.g. see Fig. 2). As we have demonstrated previously [15] the width of exchange springs can be evaluated from the reversible magnetization above $B_B$.

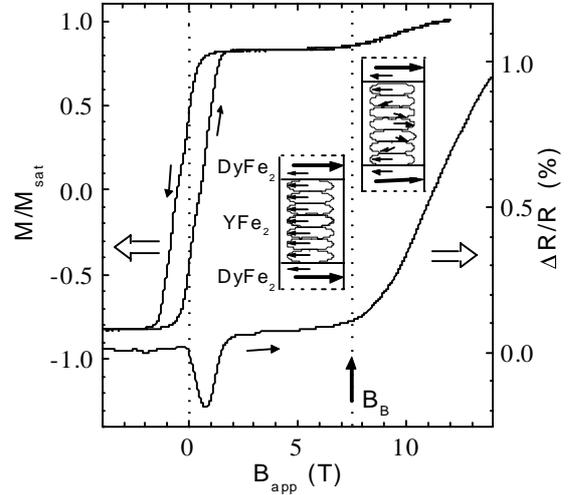

Fig. 2. Magnetization and magnetoresistance ratio $\Delta\rho/\rho$, for a current parallel to $B_{app}$, for the superlattice [60Å-$DyFe_2$/40-Å$YFe_2$]×40 at a temperature of 200 K. Insets show the spin-arrangement.

The MR for $DyFe_2/YFe_2$ samples is characterised by two distinct features one of which is associated with irreversible switching of the magnetization in the $DyFe_2$ layers at $B_C$ and the other with reversible winding/unwinding of exchange springs in the $YFe_2$ layers, for $B_{app} > B_B$. In the $DyFe_2/YFe_2$ multilayer system, both $B_C$ and $B_B$ can be varied, simply by changing the relative thicknesses of the $DyFe_2$ and $YFe_2$ layers [13, 14]. We present results for all possible cases: $B_B > B_C$ (Fig. 2), $B_B < B_C$ (Fig. 3) and $B_B \sim B_C$ (Fig. 4).

As shown in Fig. 2, in the case of $B_B > B_C$ the aforementioned two features are clearly separated. When the field is swept up from a large negative value, a dip is observed in the vicinity of the coercive field $B_C=0.7$ T. This feature is associated with AMR, similar to that seen in the DyFe$_2$ film (Fig. 1). The second feature is associated with the formation of exchange springs. The springs give rise to an increase in both the magnetization and resistivity, above $B_B =7.5$ T.

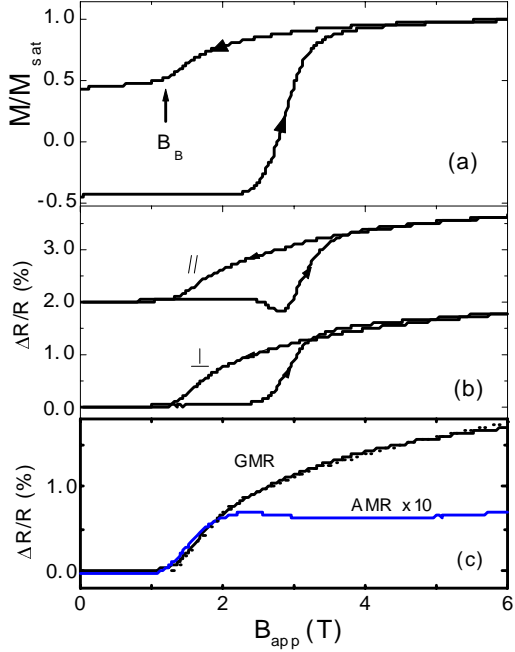

FIG. 3. Magnetization (a) and resistivity curves (b), for currents both parallel and perpendicular to $B_{app}$, for the superlattice film [150Å-DyFe$_2$/150-ÅYFe$_2$]×40 at a temperature of 100 K. The $(\Delta R/R)_{//}$ curve has been shifted up by 2%. Also shown (c) are GMR and AMR contributions (solid lines) and theoretical fit (dashed line).

To distinguish between AMR and GMR contributions we carried out measurements in two different geometries: current parallel and perpendicular to $B_{app}$. In Fig. 3(a,b), we present the magnetization, and the MR ratios $(\Delta R/R)_{\parallel}$ and $(\Delta R/R)_{\perp}$ for the superlattice [150Å-DyFe$_2$/150Å-YFe$_2$]×40 ($B_B < B_C$). On increasing the field there is a sudden upturn in both the resistances at $B_C = 2.8$ T. This is associated with the irreversible switching of the DyFe$_2$ layers, accompanied by the simultaneous creation of magnetic exchange springs in the magnetically soft YFe$_2$ layers. Note that in addition to the strong upturn, there is a small trough ~ 0.2 % in $(\Delta R/R)_{\parallel}$, again reminiscent of that seen in the DyFe$_2$ film. However, the complimentary peak in $(\Delta R/R)_{\perp}$ is masked by the strong upturn in resistivity. Secondly, on reducing the applied field back to zero, the MR follows the upper curve, eventually reaching a flat minimum at the bending field of $B_B = 1.2$ T. This part of the MR curve is fully reversible. In Fig. (3c), we show the average of ½[$(\Delta R/R)_{\perp}$+$(\Delta R/R)_{\parallel}$] for the return curves shown in Fig. 3(b). This procedure removes any AMR contributions [6]. Also shown in Fig3(c) is the AMR component calculated using ½[$(\Delta R/R)_{\perp}$-$(\Delta R/R)_{\parallel}$]. It can be seen that for the return curves the AMR effect is very weak (e.g. at 6 T the AMR contribution is 20 times smaller than the GMR). Finally, in Fig. 4, we present the MR of a [45Å-DyFe$_2$/55-YFe$_2$]×40 superlattice ($B_B \sim B_C$). A measured GMR ratio of 12 % is reached in a field of 23 T. Note that saturation has not been reached even at the highest field.

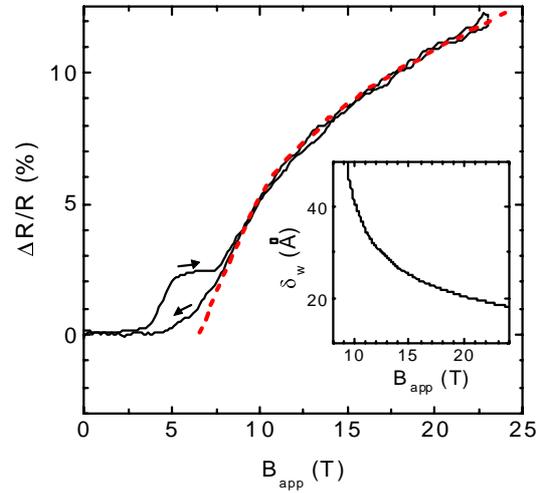

FIG. 4. Magnetoresistance ratio $\Delta\rho/\rho$ for a current perpendicular to the magnetic field $B_{app}$, for the superlattice film [45Å-DyFe$_2$/55-ÅYFe$_2$]×40 at a temperature of 100 K. The dashed line is a theoretical fit (see text). The inset shows the effective width of the exchange spring as a function of field.

For our superlattices, the MR ratio, calculated using parallel resistances, is given by:

$$\frac{\Delta R}{R} = \left[\left(\frac{t_Y}{\rho_Y} + \frac{t_D}{\rho_D}\right) \bigg/ \left(\frac{2\delta_W}{\rho_Y} - \frac{2\delta_W}{\rho_w}\right) - 1\right]^{-1} \quad (1)$$

where $\rho_Y$, $\rho_D$, $\rho_w$ are the resistivities, and $t_Y$, $t_D$, $\delta_w$ are the thicknesses of the YFe$_2$, DyFe$_2$ layers and the exchange spring, respectively. We obtained $\rho_Y$ and $\rho_D$ from measurements on thin films of DyFe$_2$ and YFe$_2$. As shown schematically in the inset to Fig. 2, the magnetic structure of an exchange spring is similar to that of a Bloch-type domain wall (see also Ref. 15). According to Levy and Zhang [7], the resistivity associated with the scattering of electrons by a domain wall in the two-channel model for the CIP geometry can be written as:

$$\rho_w = \rho_0 \left[1 + \frac{\xi^2}{5} \frac{(\rho_0^\uparrow - \rho_0^\downarrow)^2}{\rho_0^\uparrow \rho_0^\downarrow}\right] \quad (2)$$

where $\rho_0^\uparrow$ and $\rho_0^\downarrow$ are the resistivities of the spin-up and spin-down current channels, respectively and $\rho_0 = [1/\rho_0^\uparrow + 1/\rho_0^\downarrow]^{-1}$ is the resistivity without domain walls. The scaling factor $\xi$ is given by $\pi\hbar^2 k_F/4m\delta_w J$, where $J$ is the magnetic exchange constant, $k_F$ is the Fermi wave vector and $m$ is the effective mass of the electron. Consequently, in order to fit our experimental data, we need $\delta_w$ as a function of $B_{app}$. As we demonstrated in Ref. 14, the magnetic structure of the exchange springs can be extracted, using computer simulations, from the measured magnetization, associated with the exchange springs and the bending field. Here, to obtain $\delta_w$, we have calculated the thickness of a $YFe_2$ layer where the magnetic moments of Fe in the exchange spring reach a turn angle of $90^0$ and then doubled this value. The results for $\delta_w(B_{app})$ for the [45Å-$DyFe_2$/55Å-$YFe_2$]×40 superlattice are shown in the insert to Fig. 4.

To obtain $\Delta R/R = 12\%$ for the whole [45Å-$DyFe_2$/55Å-$YFe_2$]×40 multilayer structure at B=23T, we require MR due to exchange springs $\Delta\rho_w/\rho_w = 32\%$. Levy and Zhang [7] estimate that for Fe, Ni and Co, $\Delta\rho/\rho$ lies in the range 0.3% to 1.8% when $\delta_w = 150$Å. Similar figures are expected for $YFe_2$ because the $k_F$, $m$ and $J$ values are roughly similar for all these materials. According to Eq. 2 the $\Delta\rho/\rho$ ratio increases as $1/\delta_w^2$ with decreasing the domain wall width, reaching values in the range 17-100% for $\delta_w = 20$ Å, in good agreement with our result. As shown in Figs 3c and 4 the results of our calculations fit both the magnitude and the functional dependence of the MR.

The good fit to a $1/\delta_w^2$ dependence is important for understanding the origin of MR in our samples. It is known that conventional GMR multilayers consist of ferromagnetic layers separated by nonmagnetic spacers [1]. A $180^0$ exchange spring has zero average magnetic moment [11, 15], therefore an alternative way to interpret our finding would be to consider an exchange spring as a spacer, separating two ferromagnetic layers with opposite alignment of the Fe spins. However, as can be seen in Figs 2 and 3, the onset of GMR exactly coincides with the bending field. This field manifests the beginning of the rotation of Fe spins in $YFe_2$ layers. Obviously such soft exchange springs cannot be treated as nonmagnetic spacers and therefore the MR arises within the exchange springs rather than outside them. Moreover, the good agreement of our experimental results with the model by Levy and Zhang [7] over the whole magnetic field range implies that even when $180^0$ exchange springs are formed, the mechanism of GMR remains the same.

In summary, we have observed GMR in $DyFe_2/YFe_2$ superlattices which originates from magnetic exchange springs formed in the $YFe_2$ layers. Both the magnitude of the effect and its field dependence are found to be in agreement with the domain wall scattering model by Levy and Zhang [7]. The effect reaches values up to 32% in the exchange springs when their width is reduced to ~20Å.

This work has been supported by the UK Engineering and Physical Sciences Research Council, the EU Human Potential Programme (contract HPRI-1999-CT-00030), and by Technology Group 4 (Materials and Structures) of the MoD Corporate Research Programme. We also acknowledge useful discussions with G.J. Tomka.

───────────────────────────